\documentclass{llncs}
\usepackage[dvips]{graphicx}

\title{A New Approach to Efficient Enumeration by Push-out Amortization}

\author{Takeaki Uno}
\institute{
\email{uno@nii.jp}, National Institute of Informatics, 
2-1-2, Hitotsubashi, Chiyoda-ku, Tokyo 101-8430, Japan
}

\begin{document}

\maketitle

\begin{abstract}
Enumeration algorithms have been one of recent hot topics in
 theoretical computer science.
Different from other problems, enumeration has many interesting aspects,
 such as the computation time can be shorter than the total output size,
 by sophisticated ordering of output solutions.
One more example is that the recursion of the enumeration algorithm is 
 often structured well, thus we can have good amortized analysis, and 
 interesting algorithms for reducing the amortized complexity.
However, there is a lack of deep studies from these points of views; there 
 are only few results on the fundamentals of enumeration, such as 
 a basic design of an algorithm that is applicable to many problems.
In this paper, we address new approaches on the complexity analysis,
 and propose a new way of amortized analysis {\it Push Out
 Amortization} for enumeration algorithms, where the computation time
 of an iteration is amortized by using all its descendant iterations.
We clarify sufficient conditions on the enumeration algorithm
 so that the amortized analysis works.
By the amortization, we show that many elimination orderings,
 matchings in a graph, connected vertex induced subgraphs in a graph,
 and spanning trees can be enumerated in $O(1)$ time for each
 solution by simple algorithms with simple proofs.
\end{abstract}



\vspace{-10mm}
\section{Introduction}
\vspace{-2mm}

Suppose that there is a simple algorithm to solve a problem, and we 
 have two improvements on the time complexity; (a) is by developing a 
 new algorithm with a small complexity, and (b) proves that its complexity 
 is actually small by complexity analysis.
Both types of improvements are important in theoretical computer science,
 but these days almost all results are on the type of (a).
Developing simple algorithms in (a) is non-trivial, thus many recent
 algorithms and their complexity analysis are difficult to understand.
Moreover, these types of algorithms often require some structures in the
 input, hence the problem formulations tend to be distant from the real world.
On contrary, (b) type has a great advantage on these points.
Even though the analysis is complicated, we can hide the difficulty by
 producing general statements applicable to many problems.
At least, we do not have to implement the complicated proofs in a program.
According to this motivation, we study on complexity analysis in this paper,
 that is amortized analysis for enumeration algorithms.

Amortized analysis is a paradigm of complexity analysis.
In the paradigm, we charge the cost of iterations with long
 computation time to those with shorter time, to make the upper
 bound of computation time of an iteration shorter.
Compared to usual complexity analysis considering the worst case, 
 the amortized analysis is often more powerful, for example dynamic
 tree, union find, and some enumeration algorithms\cite{DynamicTree,UnionFind}.
In the case of dynamic tree, the cost of changing the shape of the tree
 is charged to the preceding changes with smaller costs, and
 attains $O(\log n)$ average time complexity for each
 change where $n$ is the size of the tree.
The time complexity is not attained by usual worst case analysis, 
 and it seems to be hard to obtain algorithms with the same
 complexity by the analysis.
This is similar to the union find algorithm, and the resulted 
 time complexity is $O(n\alpha(n))$ while straightforward algorithms
 take $O(n^2)$ time.
The concept of ``charging the cost'' made a paradigm shift on the 
 design of algorithms.
Some enumeration algorithms are designed so that the time complexity
 of an iteration is linear in the number of subproblems, to make the
 average computation time per child will be short\cite{KpRm95,SrTmUn97}.

Enumeration is now rapidly increasing its presence in theoretical 
 computer science.
One of the biggest reasons comes from its importance in application areas.
An example is the pattern mining problems in data mining.
The problem is to find all the patterns belonging to a class of structures,
 such as subsets and trees, such that the patterns satisfy some constraints
 in the given database, such as appearing at least $k$ times.
One more motivation is that there have not been many studies including
 simple problems, thus there is a great possibility.
On the other hand, enumeration has several interesting aspects which we can 
 not observe in other problems.
For example, by dealing only with the difference between output solutions,
 we can often attain the computation time shorter than its output size, by 
 outputting the solutions by the differences.
Another example is its well-structured recursion. 
We can frequently have several structural results on enumeration, and
 it gives interesting algorithms and mathematical properties, while 
 it is hard to characterize when a brunch and bound algorithm cuts off
 subproblems.
Structured recursion often gives a good amortization.
Thus, there is a great interests on investigating amortized analysis 
 on enumeration algorithms.

According to this motivation and interests, this paper addresses
 amortized analysis of enumeration algorithms.
One of our goals on this topic is to fill the gap between theory and practice.
In practice, enumeration algorithms are often quite efficient and
 than the theoretical upper bound on the computation time.
Filling the gap gives understandings for both the theoretical and 
 practical properties on the data and algorithms; the properties 
 of the data accelerating the algorithms, and the the mechanism of the
 algorithms that enable us to attain smaller bounds.

We have observed that the recursive structures of enumeration algorithms
 satisfies a property which we call {\em bottom-expanded}.
Iterations of enumeration algorithms generate several recursive calls.
Thus, the number of iterations exponentially increases 
 in deeper levels of the recursion.
On the other hand, iterations on deeper levels often have relatively
 small inputs compared to upper levels.
Thus, we can expect that iterations near by the root of the recursion are
 few and spend a long time, and iterations near by the bottom of the
 recursions are many and spend very short time.
In practice, we can frequently observe this, especially in many kinds of
 pattern mining algorithms.
This also implies that the amortized computation time per
 iteration, or even per solution, is short.
This mechanism is what we call bottom-expanded.
We can see this mechanism not only in practice but also classic
 enumeration algorithms.

This mechanism motivated us to develop a good amortized analysis.
However, amortization is not easy in general, since it is hard to globally
 estimate the number of iterations and computation time.
Thus, in many existing studies, the computation time is amortized between 
 a parent and its children, and sometimes its
 grandchildren\cite{enumPEO,Ep90,Ep94,ksubtree,KpRm95,SrTmUn97}.
These local structures are easier to analyze than the global structures.
Extensions of this idea to more global structures are non-trivial.
For example, if we want to amortize between iterations in different subtrees
 of the recursion, we have to understand the relation and the correspondence
 between all iterations in the different subtrees.
This is often a difficult task.

In this paper, we propose a new way of carrying out amortized analysis of
 the time complexity of enumeration algorithms, and propose new algorithms
 for enumeration of matchings, elimination orderings, and connected 
 vertex induced subgraphs.
We also show that the amortized analysis can prove the existing complexity
 results in very simple ways, for the enumerations of spanning trees,
 perfect elimination orderings, and perfect sequences, while the existing
  algorithms often need sophisticated algorithms or data structures.
We can also see that the condition in the analysis is often satisfied
 in practice, thus this amortized analysis explains why the enumeration
 algorithms are efficient in practice.
These satisfy out basic motivations for this kind of studies.

Our amortization of an iteration is basically done with all its descendants.
For each iteration, we push out its computation time to its children 
 so that the assigned time is proportional to their computation time.
By applying this push-out from the root of the recursion to deeper levels,
 the long computation time near the root is diffused to deeper levels,
 that have shorter time on average.
Since it is very hard to capture the structure of the recursion, we give a
 condition called {\em Push-out condition} such that the amortized
 computation time is bounded when the condition is satisfied.
As the condition is given to the relation between each iteration and its
 children, proving the satisfiability of the condition is often not difficult.

As a result, to give a bound to amortized time complexity, what we have
 to do is to prove that the condition holds for some algorithms.
In this way, we propose algorithms for enumerating matchings,
 elimination orderings, and connected vertex induced subgraphs, and 
 prove that the condition holds for each.
These lead that these graph objects can be enumerated in constant time per
 solution.
We also show that the condition holds for the algorithm for spanning tree
 enumeration, and this gives a very simple proof compared to the existing
 ones.

The paper is organized as follows.
Section \ref{sec:prlm} is for preliminaries, and Section \ref{sec:PO} describes
 our Push out amortization and Push out condition.
Sections \ref{sec:elim}, \ref{sec:match}, \ref{sec:CIS} and \ref{sec:sptree}
 show algorithms and their proofs.
We conclude the paper in Section \ref{sec:cncl}.

\vspace{-2mm}
\section{Preliminaries}\label{sec:prlm}
\vspace{-2mm}

Let $\cal A$ be an enumeration algorithm.
Suppose that $\cal A$ is a recursive type algorithm, i.e., composed of 
 a subroutine that recursively calls itself several times (or none).
Thus, the recursion structure of the algorithm forms a tree.
We call the subroutine, or the execution of the subroutine an
 {\em iteration}.
Note that an iteration does not include the computation done in the 
 subroutines recursively called by the iteration, thus no iteration is
 included in another.
When the algorithm is composed of several kinds of subroutines and
 operations, and thus the recursion is a nest of several kind of subroutines.
In such cases, we consider a series of iterations of different types as 
 an iteration.

When an iteration $X$ recursively calls an iteration $Y$, $X$ is called 
the {\em parent} of $Y$, and $Y$ is called a {\em child} of $X$.
The {\em root iteration} is that with no parent.
For non-root iteration $X$, its parent is unique, and is denoted by $P(X)$.
The set of the children of $X$ is denoted by $C(X)$.
The parent-child relation between iterations forms a tree structure
 called a {\em recursion tree}.
An iteration is called a {\em leaf iteration} if it has no child, and 
 an {\em inner iteration} otherwise.

For iteration $X$, an upper bound of the
 execution time (the number of operations) of $X$ is denoted by $T(X)$.
Here we exclude the computation for the output process from the
 computation time.
We remind that $T(X)$ is the time for local execution time, and thus does
 not included the computation time in the recursive calls generated by $X$.
For example, when $T(X) = O(n^2)$, $T(X)$ is written as $cn^2$ for some 
 constant $c$.
$T^*$ is the maximum $T(X)$ among all leaf iterations $X$.
Here, $T^*$ can be either constant, or a polynomial of the input size.
If $X$ is an inner iteration, let $\overline{T}(X) = \sum_{Y \in C(X)} T(Y)$.

In this paper, we assume that a graph is stored in a style of adjacency list.
For a vertex subset $U$ of a graph $G=(V,E)$, the {\em induced
 subgraph} of $U$ is the graph whose vertex set is $U$, and whose edge
 set contains the edges of $E$ connecting two vertices of $U$.
An edge is called a {\em bridge} if its removal increases the number
 of connected components in the graph.
An edge $f$ is said to be {\em parallel} to $e$ if $e$ and $f$ have the same
 endpoints, and be {\em series} to $e$
  if $e$ is a bridge in $G\setminus f$ and not so in $G$.

For an edge $e$ of a graph $G$, we denote the graph obtained by removing
 $e$ from $G$ by $G\setminus e$, and that by removing $e$ and edges adjacent
 to $e$ by $G^+(e)$.
Similarly, for a vertex $v$ of $G$, $G\setminus v$ is the graph obtained 
 from $G$ by removing $v$ and edges incident to $v$.
For an edge $(u,v)$ of $G$, the graph {\em contracted} by $(u,v)$, denoted 
 by $G/(u,v)$, is the graph obtained by unifying
  the vertices $u$ and $v$ into one.
For an edge set $F=\{e_1,\ldots,e_k\}$, $G/F$ denotes the graph 
 $G/e_1/e_2/\cdots/e_k$.

\vspace{-2mm}
\section{Push Out Amortization}\label{sec:PO}
\vspace{-2mm}

The size of the input of each iteration for a recursive algorithm often
 decreases as the depth of the recursion. 
Thus, iterations near the root iteration take a relatively long
 time, and iterations near leaf iterations take a relatively short time.
Motivated by this observation, we amortize the computation time by 
 moving the computation time of each iteration to its children.
We carry out this move from the top to the bottom, so that the computation
 time of ancestors is recursively diffused to their descendants.
When we can obtain a short amortized computation time in this way, iterations
 with long computation times have many descendants at least proportional
 to their computation time; the average computation time per iteration
 will be long only when they have few descendants.
However, it is not easy to prove that any inner iteration has sufficiently
 many descendants.
Instead of that, we use some local conditions, related to a parent and
 children.
Suppose that $\alpha > 1$ and $\beta\geq 0$ are two constants.\\

\vspace{-1mm}
\noindent 
{\bf \em Push Out Condition (PO condition):} for iteration $X$,
$\overline{T}(X) \ge \alpha T(X) - \beta (|C(X)|+1)T^*$.\\
\vspace{-1mm}

\begin{figure}[t]
\begin{center}
\begin{minipage}{160pt}
\vspace{-4mm}
  \begin{center}
  \includegraphics[scale=0.4]{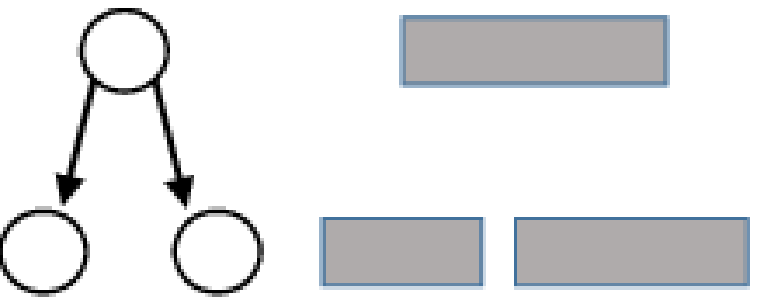}
  \end{center}
  \caption{An iteration, its children, and their computation time represented
    by rectangle lengths; seems to be inefficient if children take long time,
     but this results in many descendants indeed.
 }\label{fig:POcond}
\end{minipage}
\hspace{5mm}
\begin{minipage}{160pt}
\vspace{-4mm}
  \begin{center}
  \includegraphics[scale=0.4]{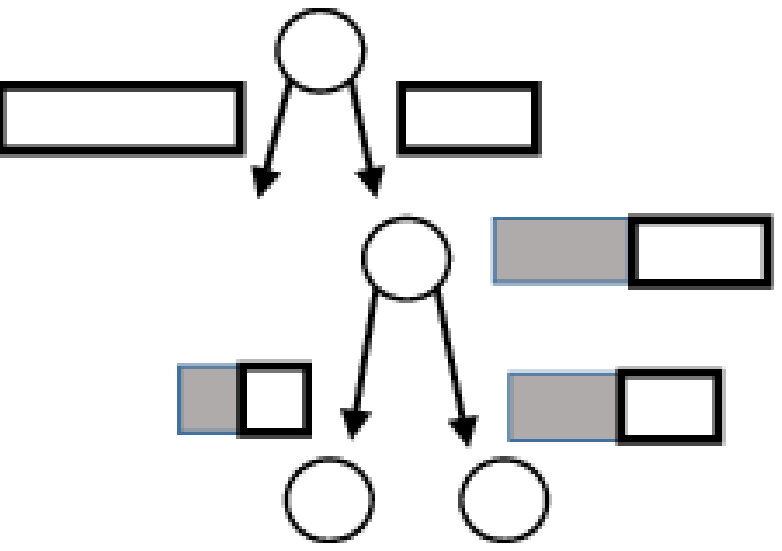}
  \end{center}
\vspace{-4mm}
  \caption{Push out rule; an iteration (center) receives computation time 
  from its parent (while rectangle), and delivers it together with its
   computation time (gray rectangle) to its children, proportional to
    their computation time.
  }\label{fig:POrule}
\end{minipage}
\end{center}
\vspace{-4mm}
\end{figure}

\noindent
Fig. \ref{fig:POcond} shows a simple example of this condition.
After the assignment of the computation time of $\alpha\beta (|C(X)|+1)T^*$
 to children and the remaining to itself, the inequation
 $\overline{T}(X) \ge \alpha T(X)$ holds.
This implies that the computation time of one level of
 recursion intuitively increases as the depth, unless there are not so many
  leaf iterations.
Considering that enumeration algorithms usually spend less time in deeper 
 levels of the recursion, we can see that this implies that each iteration
  has many children on average.
This is in some sense not a typical condition to bound the time complexity 
 of recursive algorithms; usually we want to decrease the total computation
 time in deeper levels.
However, in the enumeration, the number of leaf iterations is fixed, and 
 thereby the total computation time in the bottom level is also fixed.
Thus, this condition implies that the total computation time is short.

\begin{theorem}\label{poa}
If any inner iteration of an enumeration algorithm satisfies PO condition,
 the amortized computation time of an iteration is $O(T^*)$.
\end{theorem}
\vspace{-3mm}

\proof
To prove the lemma, we charge the computation time.
We neither move the operations nor modify the algorithm, but just
 charge the computation time; the computation time can be considered
 as tokens, and we move the tokens so that each iteration has a small
 number of tokens.
We charge the computation time from an iteration to its children, i.e.,
 from the top of the recursion tree to the bottom.
Thus, an iteration receives computation time from its parent.
We charge (push out) its computation time and that received from its parent 
 to its children.
The computation time is charged to the children, in proportion of their
 individual computation time, using the following rule.\\

\vspace{-1mm}
\noindent
{\bf \em Push out rule:}
Suppose that iteration $X$ receives a computation time of $S(X)$ from
 its parent, thus $X$ has computation time of $S(X) + T(X)$ in total.
Then, we fix $\frac{\beta}{\alpha-1}(|C(X)|+1) T^*$ of the computation
 time to $X$, and charge (push out) the remaining computation time of quantity
 $S(X) + T(X) - \frac{\beta}{\alpha-1}(|C(X)|+1) T^*$ to its children.
Each child $Z$ of $X$ receives computation time proportional to $T(Z)$,
 i.e., 

\vspace{-1mm}
\[ S(Z) = (S(X) + T(X) - \frac{\beta}{\alpha-1} (|C(X)|+1)T^*)
 \frac{T(Z)}{\overline{T}(X)}. \]
\vspace{-3mm}

\noindent
See Fig. \ref{fig:POrule} as an example.
According to this rule, we charge the computation time from the root
 iteration to leaf iterations, so that each inner iteration has 
 $O((|C(X)|+1)T^*)$ computation time.
Since the sum of the number of children over all nodes in a tree is
 no greater than the number of nodes in a tree,
  this is equivalent to that each iteration has $O(T^*)$ time.
The remaining issue is to prove the statement of the lemma by showing
 that each leaf iteration receives computation time of $O(T^*)$, 
 and it is sufficient to prove the statement.
To show that, we state the following claim.\\
 
\vspace{-2mm}
\noindent 
{\bf \em Claim}: if we charge computation time in the manner of the push
 out rule, each iteration $X$ receives computation time of at most
 $T(X) / (\alpha-1)$ from its parent, i.e., $S(X) \le T(X) / (\alpha-1)$\\
\vspace{-2mm}

\noindent
The root iteration satisfies this condition.
Suppose that an iteration $X$ satisfies it.
Then, for any child $Z$ of $X$, $Z$ receives computation time of 

\vspace{-4mm}
\begin{eqnarray*}
 && (S(X) + T(X) - \frac{\beta}{\alpha-1} (|C(X)|+1)T^*)
      \frac{T(Z)}{\overline{T}(X)}\\
 &\le& (T(X) / (\alpha-1) + T(X) - \frac{\beta}{\alpha-1} (|C(X)|+1)T^*)
      \frac{T(Z)}{\overline{T}(X)}\\
 &=& \frac{\alpha T(X) - \beta (|C(X)|+1)T^*}{\alpha-1} \times
      \frac{T(Z)}{\overline{T}(X)}\\
 &=& \frac{\alpha T(X) - \beta (|C(X)|+1)T^*}{\overline{T}(X)} \times
      \frac{T(Z)}{\alpha-1}.
\end{eqnarray*}
\vspace{-3mm}

\noindent
Since PO condition is satisfied, 
 $\overline{T}(X) \ge \alpha T(X) - \beta (|C(X)|+1)T^*$.
Thus, 

\vspace{-1mm}
\[ \frac{\alpha T(X) - \beta (|C(X)|+1)T^*}{\overline{T}(X)}
 \frac{T(Z)}{\alpha-1} \le \frac{T(Z)}{\alpha-1}. \]
\vspace{-2mm}

\noindent
By induction, any iteration satisfies the condition in the claim.
%
\qed

Note that PO condition does not require for the iterations to have
 at least two children.

\vspace{-2mm}
\section{Enumeration of Elimination Ordering}\label{sec:elim}
\vspace{-2mm}

Let ${\cal L}$ be a class of structures such as sets, graphs, and sequences.
Suppose that any structure $Z\in {\cal L}$ consists of a set of elements
 called an {\em ground set}, that is denoted by $V(Z)$.
Examples of ground sets are the vertex set of a graph, the edge set of 
 a graph, the cells of a matrix, and the letters of a string.
The empty structure $\perp$ is the unique structure that has
 $V(\perp) = \emptyset$, and hereafter we consider only $\cal L$ including 
 the empty structure.
For each $Z\in {\cal L}, Z\ne \perp$, we define the set of 
 {\em removable elements} $R(Z)$, such that for each removable element
 $e\in R(Z)$, the removal of $e$ from $Z$ results in a structure
 $Z'\in {\cal L}, V(Z') = V(Z)\setminus \{ e\}$.
We denote the removal of $e$ from $Z$ by $Z\setminus e$, and we assume that
 no two different structures can be generated by the removal of $e$.
By using removable elements, we define {\em elimination orderings}.
An elimination ordering is an ordering $(z_1,\ldots,z_n)$ of
 elements in $V(Z)$ iteratively removed from $Z$ until $Z$ is $\perp$, i.e.,
 any $z_i$ is removable in the structure $Z_i$ that is obtained by repeatedly 
 removing $z_1$ to $z_{i-1}$ from $Z$.
Example of elimination ordering are removing leaves from a tree,
 and perfect elimination ordering of a chordal graph.
A simple algorithm for enumerating elimination orderings can be described
 as follows.

\begin{tabbing}
{\bf Algorithm} EnumElimOrdering ($Z, S$)\\
1. {\bf if} $|V(Z)| = 1$, {\bf output} $S + z$ where $V(Z) = \{ z\}$;
 {\bf return}\\ 
2. {\bf for} each element $z\in V(Z)$ {\bf do}\\
3. \ \ {\bf if} $z\in R(Z)$, {\bf call} EnumElimOrdering
 ($Z\setminus z, S + z$)\\
4. {\bf end for}
\end{tabbing}

Suppose that we are given a structure $Z$ in a class $\cal L$ and
 removable ground set $R$ for ground set $V(Z)$.
We suppose that for any $z\in V(Z)$, we can list all $z\in R(Z)$
 in $\Theta(p(|V(Z)|)q(n))$ time, where $p(|V(Z)|)$ is a polynomial
 of $|V(Z)|$, and $q(n)$ is a function where $n$ is an invariant of the
 input structure, such as the number of edges in the original graph.
We also assume that a removal of element takes $\Theta(p(|V(Z)|)q(n))$ time.

\begin{theorem}\label{elim}
Elimination orderings of a class $\cal L$ can be enumerated
 in $O(q(n))$ time for each, if $|R(Z)| \ge 2$ holds for each $Z\in {\cal L}$
 such that $|V(Z)|$ is larger than a constant number $c$.
\end{theorem}

\proof
We first bound the computation time except for the output processes, 
 that is, step 1 of EnumElimOrdering.
First, we choose two constants $\delta>c$ and $\alpha>1$ such that 
 $\frac{2p(i-1)}{p(i)} > \alpha$ holds for any $i > \delta$.
Since $p$ is a polynomial function, $\frac{p(i)}{p(i-1)}$ converges to
 $1$, thus such $\alpha$ always exists.
Let $X$ be an iteration.
When $X$ inputs $Z$ with $|V(Z)| \le \delta$, the computation time 
 is $q(n)$, except for the output process.
Hence, we have $T^* = O(q(n))$.
For the case $|V(Z)| \le \delta$, the computation time of $X$ is bounded
 by $q(n)$.
For the case $|V(Z)| > \delta$, we have 

\vspace{-2mm}
\[ \overline{T}(X) \ \ge \ 2(|V(Z)|-1)p(|V(Z)|-1)q(n)
 \ > \ \alpha |V(Z)|p(|V(Z)|)q(n),\]
\vspace{-3mm}

\noindent
since $X$ has at least two children.
Thus, $X$ satisfies PO condition with any constant $\beta > 0$.
From Theorem \ref{poa}, except for the output process, the computation
 time is bounded by $O(q(n))$ time for each iteration whose input
  has at least $\delta$ elements.
Since any inner iteration $Y$ has exactly one child only if $|V(Y)| \le c$,
 the number of inner iterations is bounded by the number of leaf
 iterations, multiplied by $c$.
Therefore, the computation time for each elimination ordering can be 
 bounded by $O(cq(n)) = O(q(n))$ time.

Next, let us consider the output process.
Instead of explicitly outputting elimination orderings, we output
 each elimination ordering $S$ by the difference from $S'$ that is
  output just before $S$.
We can output them compactly in this way.
Although the difference can be large up to $|V(Z)|$, we can see
 that it is bounded by the number of operations done from the previous 
 output process.
Thus, the size of all output differences, except for the first one output 
 in the usual way, is at most proportional to the total computation time.
Therefore, the computation time for the output process
 is also bounded by $O(q(n))$ time for each.
\qed

The next corollary immediately follows from the theorem.

\begin{corollary}
For a given set class, elimination ordering can be enumerated 
 by EnumElimOrdering in $O(1)$ amortized time for each, if each
 inner iteration generates at least two recursive calls, and 
 takes $O(p(|V(Z)|))$ time, where $p$ is a polynomial of $|V(Z)|$. \qed
\end{corollary}

There are actually several elimination orderings to which this 
 theorem can be applied, and they are listed below.
For conciseness, we have described each by their structures and
 removable elements.\\

\vspace{-1mm}
{\bf Example (a): perfect elimination orderings of a chordal
 graph\cite{enumPEO}}\\
For a graph, a vertex is called {\em simplicial} if the vertices adjacent
 to it form a clique.
An elimination orderings of simplicial vertex is called {\em perfect
elimination ordering}\cite{elimord}, and a graph is {\em chordal}
 if it has a perfect elimination ordering.
We define $\cal L$ by the set of chordal graphs, $V(Z)$ by the vertex
 set of $Z\in {\cal L}$, and $R(Z)$ by the set of its simplicial vertices.

It is known that any chordal graph $Z$ admits a clique tree whose
 vertices are maximal cliques of $Z$.
If $Z$ is a clique, all vertices in $Z$ are simplicial.
If not, it is known that there are at least two cliques that has a vertex
 that is not included in the other maximal cliques.
Note that these cliques are leaf cliques of a clique tree, where 
 the vertices of a clique tree are maximal cliques of $Z$, 
 each edge connects overlapping cliques, and the maximal cliques 
 including any vertex forms a subtree of the clique tree.
The vertex is simplicial, hence $|R(Z)| \ge 2$ always holds.
Since we can check whether a vertex is simplicial or not in $(|V(X)|^2)$
 time, we can enumerate all perfect elimination orderings in $O(1)$
 time for each.
Note that although the algorithm in \cite{enumPEO} already attained the same 
 time complexity, our analysis yields much simpler algorithm and proof,
\\

\vspace{-1mm}
{\bf Example (b): perfect sequence\cite{perfectseq}}\\
$\cal L$ is the class of chordal graphs $Z$, and $V(Z)$ is the set of
 maximal cliques in $Z$.
A maximal clique is removable if it is a leaf of some clique trees of $Z$,
 and the removal of a maximal clique $z$ from $Z$ is the removal of all
  vertices of $z$ that do not belong to another maximal clique.
The removal of the vertices results in the graph that includes remaining 
 maximal cliques, and no new maximal clique appears in the graph.
Note that a clique tree has at least two leaves if it has more than one 
 vertex, thus $|R(Z)|\ge 2$.
An elimination ordering is called a {\em perfect sequence}.
Since all removable maximal cliques can be found in polynomial
 time in the number of maximal cliques\cite{perfectseq},
 all perfect sequences are enumerated in $O(1)$ time for each.\\

\vspace{-1mm}
The elimination orderings induced by following removable elements 
 can be also enumerated in $O(1)$ time for each.
 
\vspace{-2mm}
\begin{itemize}
\item non-cut vertices of connected graph
\item points on surface of convex hull of a point set in plane
\item leaves of a tree
\item vertices of degrees less than seven of a simple planar graph.
\end{itemize}

\vspace{-4mm}
\section{Enumeration of Matchings}\label{sec:match}
\vspace{-2mm}

A {\it matching} of a graph is an edge subset of a graph $G=(V,E)$
 such that no two edges are adjacent.
The matchings are enumerated by the following algorithm.

\begin{tabbing}
{\bf Algorithm} EnumMatching ($G=(V,E), M$)\\
1: {\bf if} $E = \emptyset$ {\bf then output} $M$; {\bf return}\\
2: choose an edge $e$ from $E$\\
3: {\bf call} EnumMatching ($G\setminus e, M$)\\
4: {\bf call} EnumMatching ($G^+(e), M\cup \{ e\}$)
\end{tabbing}

\noindent
The time complexity of an iteration of EnumMatching is $O(|V|)$.
Since each inner iteration generates two children, the computation
 time for each matching is $O(|V|)$, and no better algorithm
 has been proposed in the literature.
A leaf iteration takes $O(1)$ time, thus $T^* = O(1)$.
However, PO condition may not hold for some iterations.
This cannot be better than $O(|V|)$ in straightforward ways.

PO condition does not hold when many edges are adjacent to $e$.
In such cases, $G^+(e)$ has few edges, thus the subproblem of
 $G^+(e)$ takes short time so that PO condition does not hold.
To avoid this situation, we modify the way of recursion as follows
 so that in such cases the iteration has many children.
Let $u_1,\ldots,u_k$ be the vertices adjacent to $v$, and $e_i = (v,u_i)$.
We partition the matchings to be enumerated into 

\vspace{-1mm}
\begin{itemize}
\item matchings including $e_1$
\item matchings including $e_2$
\item $\cdots$
\item matchings including $e_k$
\item matchings including no edge incident to $v$.
\end{itemize}
\vspace{-1mm}

We see that any matching belongs to exactly one of these groups.
To recur, we derive $G^+(e_1),\ldots,G^+(e_k)$ and $G\setminus v$.
$G\setminus v$ and $G^+(e_1)$ can be derived in $O(|E|)$ time.
To shorten the computation time for $G^+(e_i)$ for $i\ge 2$, 
 we construct $G^+(e_i)$ from $G^+(e_{i-1})$.
We add all edges of $G$ incident to $u_{i-1}$ to $G^+(e_{i-1})$,
 and remove all edges adjacent to $u_i$, and obtain $G^+(e_i)$.
This can be done in $O(d(u_{i-1})+d(u_i))$ time.
To construct $G^+(e_i)$ for all $i=2,\ldots,k$, we need 

\vspace{-1mm}
\[ O(\ (d(u_1)+d(u_2))\ +\ (d(u_2)+d(u_3))\ +\ \cdots\ +\ 
 (d(u_{k-1})+d(u_k))\ )\ \ =\ O(|E|) \]
\vspace{-3mm}

\noindent
time.
Thus, the computation time of an iteration is bounded by $c|E|$
 with a constant $c$.
The algorithm is described as follows.

\begin{tabbing}
{\bf Algorithm} EnumMatching2 ($G=(V,E), M$)\\
1: {\bf if} $E = \emptyset$ {\bf then output} $M$; {\bf return}\\
2: choose a vertex $v$ having the maximum degree in $G$\\
3: {\bf call} EnumMatching2 ($G\setminus v, M$)\\
4: {\bf for} each edge $e$ adjacent to $v$,
 {\bf call} EnumMatching2 ($G^+(e), M\cup \{ e\}$)
\end{tabbing}

\begin{theorem}\label{match}
All matchings in a graph can be enumerated in $O(1)$ time for each,
 with $O(|E|+|V|)$ space.
\end{theorem}

\proof
The amortized computation time for outputting process is
 bounded by $O(1)$ for each by using difference as elimination ordering.
Let us consider an inner iteration $X$.
In the iteration $X$, if $d(v)\ge |E|/4$, we generate at least $|E|/4$
 recursive calls, thus we have $|C(X)|=\Omega(|E|)$ and PO
 condition is satisfied by choosing sufficiently large $\beta$.
If $d(v) < |E|/4$, the subproblems of $G\setminus v$ take 
 at least $\Theta(3c|E|/4)$ time, and the subproblems of $G^+(e_1)$
 take at least $c|E|/2$ time.
Hence, by setting $\alpha = 1.25$, we have 

\vspace{-1mm}
\[ \overline{T}(X) \ge 3c|E|/4 + c|E|/2 = 5c|E|/4 \ge \alpha T(X) - \beta |C(X)|T^* \] 
\vspace{-3mm}

\noindent
 thereby PO condition holds.
Remind that each inner iteration generates two or more recursive calls, 
 the number of iterations does not exceed the twice the number of matchings.
Since any inner iteration satisfies PO condition and
 $T^* = O(1)$, the statement holds.
We remind that we assumed that there is no isolated vertex in the
 input graph, and thus the number of matchings in the graph is 
 greater than the number of vertices, and the number of edges.
\qed

\vspace{-2mm}
\section{Enumeration of Connected Vertex Induced Subgraphs}\label{sec:CIS}
\vspace{-2mm}

We consider the problem of enumerating all vertex sets of the given graph $G=(V,E)$ inducing connected subgraphs (connected induced subgraphs in short).
In literature, an algorithm is proposed that runs in $O(|V|)$ time for
 each\cite{AvFk96}.
For the enumeration, it is sufficient to enumerate all connected 
 induced subgraphs including the given vertex $r$.
For a vertex $v$ adjacent to $r$, the connected induced subgraphs
 including $r$ are partitioned into those including $v$ and
 those not including $v$.
The former subgraphs are connected induced subgraphs in 
 $\underline{G/(r,v)}$ and the latter subgraphs are those in $G\setminus v$.
We have the following algorithm according to this partition, and we 
 prove that this algorithm satisfies PO condition.

\begin{tabbing}
{\bf Algorithm} EnumConnect ($G=(V,E), S, r$)\\
1: {\bf if} $d(r) = 0$ {\bf then output} $S$; {\bf return}\\
2: choose a vertex $v$ adjacent to $r$\\
3: {\bf call} EnumConnect ($\underline{G/(r,v)}, S\cup \{ v\}, r$)\\
4: {\bf call} EnumConnect ($G\setminus v, S, r$)
\end{tabbing}

\begin{theorem}\label{connect}
All connected vertex induced subgraphs in a graph can be enumerated in $O(1)$
 time for each, with $O(|E|+|V|)$ space.
\end{theorem}

\proof
The correctness of the algorithm and the bound for memory usage are clear.
Since each inner iteration generates exactly two recursive calls, 
 the number of iterations is linearly bounded by the number of connected
 induced subgraphs, and $T^* = O(1)$.

As same to the matching enumeration, the computation time for outputting 
process is bounded by $O(1)$ for each.
An inner iteration $X$ of the algorithm takes $O(d(r)+d(v))$ time.
We assume that $T(X) = c(3d(r)+d(v))$ for a constant $c$, and 
 leaf iteration takes $3c$ time, since $T^* = O(1)$.
The constant factor of three is a key to PO condition.

The degree of $r$ is at least $(d(r)+d(v))/2-1$ in $\underline{G/(r,v)}$, and 
 $d(r)-1$ in $G\setminus v$.
Note that $d(r)$ and $d(v)$ are degrees of $r$ and $v$ in $G$.
From this, we can see that the child iteration of $\underline{G/(r,v)}$ takes
 at least $3c((d(r)+d(v))/2-1)$ time, and that of $G\setminus v$ takes
 at least $3c(d(r)-1)$ time.
Their sum is at least

\vspace{-1mm}
\[ 3c((d(r)+d(v))/2-1) + 3c(d(r)-1) =
 \frac{3}{2}c(3d(r)+d(v)) - 6c = \frac{3}{2}T(X) - 6c.\]
\vspace{-3mm}

\noindent
Setting $\beta = 6$, we can see that $X$ satisfies PO condition.
Thanks to Theorem \ref{poa}, the computation time for each connected
 induced subgraph is $O(1)$.
\qed

\vspace{-2mm}
\section{Spanning Trees}\label{sec:sptree}
\vspace{-2mm}

A subtree $T$ of a graph $G=(V,E)$ is called a {\em spanning tree} if any 
 vertex of $G$ is incident to at least one edge of $T$.
Any spanning tree has $|V|-1$ edges.
There have already been several studies on this
 problem\cite{KpRm95,SrTmUn97,Un99}, and \cite{Un99} is the simplest
 and uses an amortized analysis similar to us.
Without loss of generality, we assume that the input graph does not
 have any bridge.

Let $e_1$ be an edge of $G$.
If there are several edges $e_2,\ldots,e_k$ parallel to $e_1$,
 let $F = \{ e_1,\ldots,e_k\}$ and $F_i = F\setminus \{ e_i\}$.
We see that at most one edge from $F$ can be included in a spanning tree, 
 thus we enumerate spanning trees in $(G\setminus F_1)/e_1,\ldots,
(G\setminus F_k)/e_k$.
We further enumerate spanning trees in $G\setminus F$ if it is connected.
Any spanning tree is enumerated in exactly one of these.
When $e_1$ has no parallel edges, $e_1$ can have series edges.
If there are several edges $e_2,\ldots,e_k$ series to $e_1$, again
 let $F = \{ e_1,\ldots,e_k\}$ and $F_i = F\setminus \{ e_i\}$.
We also see that any spanning tree includes at least $k-1$ edges of $F$,
 thus we enumerate spanning trees in $(G/F_1)\setminus e_1,\ldots, 
 (G/F_k)\setminus e_k$.
We further enumerate spanning trees in $G/F$ if $F$ is not the edges of
 a cycle.
Also in this case, any spanning tree is enumerated once among these.
By using these subdivisions, we construct the following algorithm.

\begin{tabbing}
{\bf Algorithm} EnumSpanningTree ($G=(V,E), T$)\\
1: {\bf if} $E = \emptyset$ {\bf then output} $T$; {\bf return}\\ 
2: choose an edge $e_1$ from $E$\\
3: $F^p := \{ e_1\} \cup \{ e | e \mbox{ is parallel to } e_1\}$ ;\\ 
\ \ \ \ \ $F^s := \{ e_1\} \cup \{ e| e \mbox{ is not parallel to } e_1,
   \mbox{ and } e \mbox{ is series to } e_1\}$\\
4: {\bf for} each $e_i\in F^p$, \ {\bf call} EnumSpanningTree
 ($(G\setminus (F^p\setminus \{e_i\})/ e_i, T\cup \{ e_i\}$)\\
5: {\bf for} each $e_i\in F^s$, \ {\bf call} EnumSpanningTree
 ($(G/ (F^s\setminus \{e_i \})\setminus e_i, T\cup (F^s \setminus \{ e_i\})$)
\end{tabbing}

\noindent
We observe that these $k$ subgraphs are actually isomorphic
 in both cases except for the edge label $e_i$, thus constructing
 these graphs takes $O(|V|+|E|)$ time.

\begin{theorem}\label{spantree}
All spanning trees in a graph can be enumerated in $O(1)$ time
 for each, with $O(|E|+|V|)$ space.
\end{theorem}

\proof
The space complexity of the algorithm is $O(|E|+|V|)$ and an 
iteration takes $\Theta(|V|+|E|)$ time since all edges parallel/series to
 an edge can be found by two connected
 component decomposition in $O(|V|+|E|)$ time.
If no edge is parallel or series to $e_1$,
 we generate two subproblems of $|E|-1$ edges, thus PO condition holds.
If $k$ edges are parallel or series to $e_1$, we have at least $k+1\ge 2$
 subproblems of $|E|-(k+1)$ edges.
When $k+1\ge |E|/4$, $T(X) - \beta (|C(X)|+1)T^* = 0 $ holds for some 
 $\beta>0$, and PO condition holds.
When $k+1< |E|/4$, $(k+1)(|E|-(k+1)) \ge 1.5|E|$ holds, 
 PO condition holds for $\alpha = 1.5$ and some $\beta>0$.
Since each iteration generates at least two recursive calls or outputs a
 solution, the number of iterations is at most twice the number of solutions,
 therefore the statement holds.
\qed

\vspace{-2mm}
\section{Conclusion}\label{sec:cncl}
\vspace{-2mm}

We introduced a new way of looking at amortizing the computation time of 
 enumeration algorithms, by local conditions of recursion trees.
We clarified the conditions that are sufficient to give non-trivial upper 
 bounds for the average computation time of iterations that only depended
 on the relation between the computation time of a parent iteration and
  that of its child iterations.
We showed that many algorithms for elimination orderings have good
 properties so that the conditions are satisfied, and thus enumerated 
 in constant time for each.
Several other enumeration algorithms for matchings, connected vertex
 induced subgraphs, and spanning trees were also described, whose
 time complexities are $O(1)$ for each solution.

There are many problems for those enumeration algorithms that do not
 satisfy the conditions.
An interesting future work is to develop new algorithms for these
 problems, that satisfy the conditions.
Another direction is to study other conditions for bounding 
 amortized computation time.
Further studies on amortized analysis will possibly fill the gaps between 
 theory and practice, and clarify the mechanisms of enumeration algorithms.
 
\ \\
\noindent
{\bf \large Acknowledgments: }
Part of this research is supported by the Funding Program for World-Leading
 Innovative R\&D on Science and Technology, Japan, and Grant-in-Aid for
 Scientific Research (KAKENHI), Japan.

\end{document}